\documentclass[reprint,twocolumn]{revtex4}
\usepackage{graphicx}%
\usepackage{dcolumn}%
\usepackage{bm}%
\usepackage{amssymb}
\usepackage{epsfig,colordvi}
\usepackage[latin1]{inputenc}
\usepackage{color} 

 \def\be{\begin{equation}}
 \def\ee{\end{equation}}
 \def\bea{\begin{eqnarray}}
 \def\eea{\end{eqnarray}}
 \def\bean{\begin{eqnarray*}}
 \def\eean{\end{eqnarray*}}
 \def\gsim{\mathrel{\rlap{\lower0.2em\hbox{$\sim$}}\raise0.2em\hbox{$>$}}}
 \def\ksim{\mathrel{\rlap{\lower0.2em\hbox{$\sim$}}\raise0.2em\hbox{$<$}}}
 \def\kg{\mathrel{\rlap{\lower0.25em\hbox{$>$}}\raise0.25em\hbox{$<$}}}

 \newcommand{\eq}[1]{(\ref{#1})}
\newcommand{\nuc}[2]{$^{#1}#2$}

\newcommand{\rneq}{\hbox{$R_P$=$R_N$ }}
\newcommand{\rnlt}{\hbox{$R_P\!\!<\!\!R_N$ }}
\newcommand{\rns}[1]{\hbox{$\Delta R$=#1 }}

\begin{document}

\title{The influence of the neutron skin and the asymmetry energy on the $\pi^-/\pi^+$ ratio}

\author{C. Hartnack$^1$, A. Le F\`evre$^2$, Y. Leifels$^2$, and J. Aichelin$^{1,3}$}
\affiliation{$^1$ SUBATECH, UMR 6457  IMT Atlantique, Universit\'e de Nantes, IN2P3/CNRS
\\ 4 rue Alfred Kastler, 44307 Nantes cedex 3, France}
\affiliation{$^2$ GSI Helmholtzzentrum f\"ur Schwerionenforschung GmbH,
  Planckstr. 1, 64291 Darmstadt, Germany} 
\affiliation{$^3$ Frankfurt Institute for Advanced Studies, Ruth Moufang Str. 1
\\ 60438 Frankfurt, Germany}

\date{\today}

\begin{abstract} \noindent
We use the Isospin Quantum Molecular Dynamics model (IQMD) to analyze the centrality dependence of the isospin ratio of pions, 
$\pi^-/\pi^+$. We find that the density dependence of the asymmetry potential, the Pauli blocking of the $\Delta$-decay and the thickness
of the neutron skin influence in different ways this observable. Using the centrality dependence of this ratio at different beam energies we can
disentangle the different contributions and open the way for their experimental determination. 
\end{abstract}

\pacs{12.38Mh}

\maketitle

\section{Introduction}

The density dependence of the $\pi^-/\pi^+$ multiplicity ratio as a function of beam energy, measured by
the FOPI collaboration  \cite{Reisdorf:2006ie}, has triggered a lot of
theoretical interest (see eg. \cite{Xiao:2008vm, Xiao:2013awa, Feng:2009am, Hong:2013yva,Cozma:2014yna}). 
This ratio sheds light on the asymmetry energy, one of the
hot topics in present heavy ion research at beam energies around 1 AGeV. 
Their density (and momentum dependence) is interesting in itself but also of great
importance to link the (almost) symmetric matter, studied in heavy ion collisions, with the properties of asymmetric
matter which have to be known if one wants to study astrophysical questions like the merger
of neutron stars or to understand their mass to radius relation.

Chiral perturbation theory allows for calculating reliably the asymmetry energy in nuclear
matter up to normal nuclear matter density \cite{schwenk}. Beyond that presently calculations are
not at hand and in addition there may be a strong momentum dependence. Numerical programs, which 
model the entire heavy ion collisions on a computer and which assume that the proton and neutron rms radius 
is identical, predict quite reasonable a large variety of observables 
but failed to describe the excitation function of the  $\pi^-/\pi^+$ ratio. Hong and Danielewicz \cite{ Hong:2013yva}
introduced a momentum dependent mean field which allows to reproduce  the  $\pi^-/\pi^+$ ratio
at beam energies beyond 700 AMeV . They did  not find any dependence of this ratio on the density
dependence of the asymmetry energy for beam energies at and below 400 AMeV. Feng and Jin 
\cite{ Feng:2009am} found, however, for the lowest beam energy (400 AMeV) a strong dependence 
of the  $\pi^-/\pi^+$ ratio  on the density dependence of the asymmetry energy whereas  Cozma \cite{Cozma:2014yna}
studies the influence of a isospin dependent $\Delta N$ potential on the $\pi^-/\pi^+$ ratio. 
Wei  \cite{Wei:2013sfa} addresses the influence of the neutron skin on the $\pi^-/\pi^+$ ratio and found
it  to be negligible for the $\pi^-/\pi^+$ ratio measured in central collisions. 

It is the purpose of this Letter to demonstrate that the centrality dependence of the  $\pi^-/\pi^+$ ratio 
allows on the one side to measure the neutron skin of nuclei and is, on the other side, sensitive to the
density dependence of the asymmetry energy. The asymmetry energy has no influence on this ratio for very peripheral collisions 
hence they can be used to determine the thickness of the neutron skin.
Having determined the thickness of the neutron skin from peripheral collisions we can determine 
it's influence at central collisions, which is not negligible at lower beam energies, and finally determine
the modification of the $\pi^-/\pi^+$ ratio due to  the asymmetry energy. It turns out that this is a quite robust procedure.

\section{Density profiles for protons and neutrons}

For our calculations we use the IQMD approach \cite{IQMD-first}, a microscopical transport model calculating heavy ion collisions on an event-by-event basis. Details of the model can be found in \cite{IQMD-details}. Here we only underline that in this kind of models particles are represented by Gaussian distributions and interact via two-body interactions.
For our analysis we have modified the initialization of the nuclei which had, up to now, as all other models with the exception of
\cite{Wei:2013sfa},  the same rms radius for protons and neutrons. 
In standard IQMD calculations (as in many other microscopic models) the centroids of the Gaussians are distributed inside a sphere in the rest frame of the nucleus according to
\begin{equation}
\left ( \vec{r}_i - \vec{r}_{CM} \right )^2 \le R_A^2 \qquad R_A =R_0 \cdot A^{1/3}
\label{def-RA}
\end{equation}
where $\vec{r}_i$ and $\vec{r}_{CM}$ are the position vectors of particle $i$ and of the center-of-mass of the nucleus, respectively. $R_0=1.12\quad \rm fm$ is the radius parameter and $A=Z+N$ is the number of nucleons of the nucleus.
This initialization, which we will call "\rneq", assures the same rms radius for protons and neutrons even for heavy isospin-asymmetric systems. Consequently,  in the whole nucleus the neutrons have systematically a higher density than the protons. 
If we want to assume the same density of protons and neutrons at least in the centre of the nucleus we have to allow protons and neutrons to have different rms radii, which can be obtained by the distribution of the centroids of the Gaussians according to
\begin{equation}
 R_P =R_0 \cdot (2Z)^{1/3} \qquad R_N =R_0 \cdot (2N)^{1/3}
\label{def-RPZ}
\end{equation}
where $R_P$ and $R_N$ denote the radii for protons and neutrons and which we will call "\rnlt". 
However, this initialization yields a difference of the rms radii  of protons and neutrons of around 0.5 fm in a system like $^{208}Pb$. 
The truth might probably lie in between both (rather schematic) options therefore we will also allow an initialization with an 
intermediate neutron skin thickness $s$ according to
\begin{equation}
 \Delta R =R_N -R_P =s
\label{def-RPs}
\end{equation}
which we will call "\rns{s}". The proton and neutron radii are now related to $R_A$ by  
\begin{equation}
 R_P =R_A -s/2 \qquad R_N =R_A +s/2. 
\label{def-RNRP}
\end{equation}
A typical value for $s$ might be around  $s=0.2 \to 0.3$ fm for  $^{208}Pb$. 
It should be noted that even if the distribution of the centroids according to equation \eq{def-RA} seems to be a sharp function, the Gaussian-like density distribution of each nucleon produces a quite smooth density distribution of protons and neutrons. 

\begin{figure}[htb]
\includegraphics[width=0.9\columnwidth]{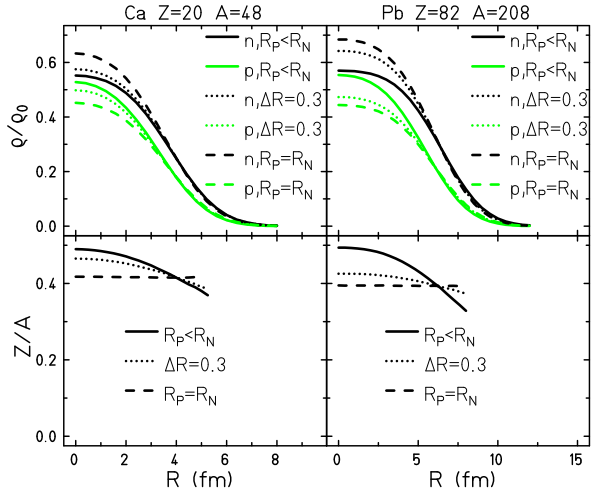} 
\caption{Top: Density profiles for protons (green lines) and neutrons (black lines) for different types of initialization. 
Left: for a \nuc{48}{Ca} nucleus, right: for a \nuc{208}{Pb} nucleus.
Bottom: charge distribution $Z/A(R)$.}
\label{rho-profiles}
\end{figure}

This can be seen in Fig. \ref{rho-profiles}, which presents in the top row the density distributions of protons (green lines) 
and neutrons (black lines) for a \nuc{48}{Ca} nucleus (left) and for a \nuc{208}{Pb} nucleus (right).
Different options have been used: the initialisation using equation \eq{def-RA} only (\rneq) is represented by dashed lines, 
while the initialization using \eq{def-RPZ}  (\rnlt) is given by full lines and an initialization using a neutron skin thickness of 0.3 fm, 
as shown in \eq{def-RPs},  (\rns{0.3}) is marked by dotted lines. 
A first glance might suggest that the initialization only effect the inner part of the nucleus (small values of $R$) where indeed the 
difference in densities is most significant. Moreover, for radii close to $R_A$ (4 fm for \nuc{48}{Ca} and 6.6 fm for 
\nuc{208}{Pb}) the difference between neutrons and protons seems even to be independent of the chosen initialization.
However, if we now look on the charge distributions ($Z/A(R)$, bottom row) we see a completely different behavior of the initialization: 
the initialization using equation \eq{def-RA} only (\rneq, dashed line) yields an overall constant value for $Z/A$, 
while the initialization using \eq{def-RPZ}  (\rnlt, full line) starts isospin-equilibrated ($Z/A=0.5$) in the center 
and falls down strongly in the peripheral region. The initialization using a neutron skin thickness of 0.3 fm, as shown in \eq{def-RPs},  
(\rns{0.3}, dotted line) shows a behavior lying in between both other options.

\section{Isospin ratios of pions}

These different $Z/A(R)$ ratios have to be kept in mind when we study now the production of pions. The pion production in IQMD is done via the $\Delta$-channel, where $\Delta$s can be produced in nucleon-nucleon ($NN$) collisions and be reabsorbed in $N\Delta$ collisions. The $\Delta$ decays and produces a free pion, which can be reabsorbed in collisions with a nucleon to form a $\Delta$ again:
\be
NN \leftrightarrow N \Delta \qquad \Delta \leftrightarrow N \pi.
\ee
These reactions have to comply with detailed balance and isospin effects have to be taken into account by the use of Clebsch-Gordon coefficients. For more details see \cite{IQMD-pions}.

\begin{figure}[htb]
\includegraphics[width=0.9\columnwidth]{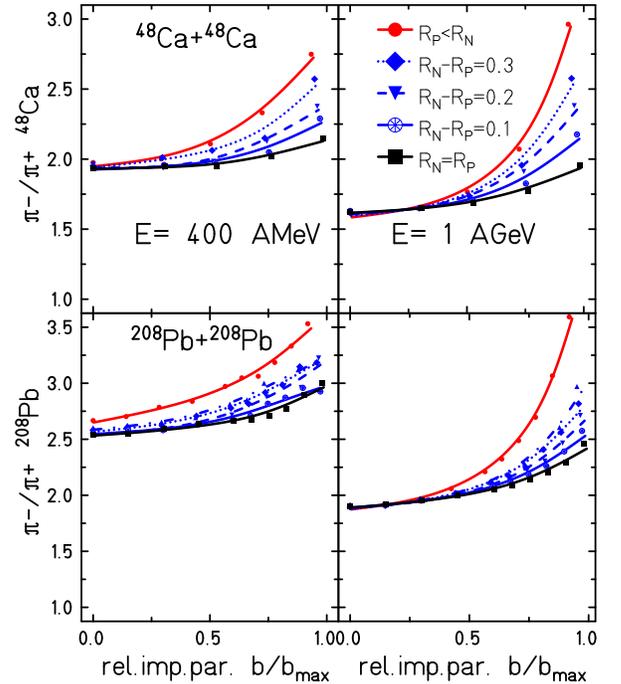} 
\caption{The isospin ratio of pions $\pi^-/\pi^+$ as a function of the impact parameter $b$ for different initialization. 
Left: for an incident energy of 400 AMeV, right: for 1000 AMeV, 
upper row for \nuc{48}{Ca}, lower row for \nuc{208}{Pb}.  }
\label{ratios-profiles}
\end{figure}

The effect of the different initialization procedures (and the corresponding different isospin ratios of the nucleons) on the isospin ratio of pions
as a function of centrality, $\pi^-/\pi^+ (b)$, can be depicted from Fig. \ref{ratios-profiles} which presents results for the symmetric 
systems \nuc{48}{Ca}+\nuc{48}{Ca} (top row) and \nuc{208}{Pb}+\nuc{208}{Pb} (bottom row), 
both for an incident energy of 400 MeV/nucleon (left hand side) and 1 GeV/nucleon (right hand side).
While the initialization using equal radii for protons and nucleons (\rneq, black line with squares) 
shows a quite moderate dependence of the isospin ratio on the centrality, the initialization assuring the same
density for protons and neutrons in the center (and thus largely different radii, \rnlt, red line with bullets) shows a strong 
enhancement of that ratio at very peripheral collisions. The initialization with intermediate assumptions on the neutrons skin 
(\rns{0.1 \ldots 0.4}, blue lines) show some intermediate enhancement of that ratio at peripheral collisions. 
The overall values of the ratios depend on the size of the system: The larger \nuc{208}{Pb} system yields higher ratios than \nuc{48}{Ca} for central and for peripheral collisions. It should be noted, that both systems do not differ very much in the maximum density reached during the collision but strongly in the number of  rescatterings of pions. It should also be noted that the enhancement at peripheral collisions becomes more significant when rising the incident energy from 400 AMeV to 1 AGeV.

Let us first discuss the right hand side corresponding to the incident energy of 1 GeV/nucleon. We see that for both systems the ratio becomes independent of the chosen initialization profile when going to very central collisions. This can be understood if we assume that there is enough energy to produce pions anywhere in the nuclei. In this case the individual $Z/A(R)$ profiles do not effect the total yields, which are only determined by the overall $Z/A$ in the system, a value only depending on the charge and mass of both nuclei. It should be noted that a simple combinatorial estimation of pion yields, assuming that each nucleon can collide with each other nucleon and taking into account the isospin Clebsch-Gordon coefficients for $\Delta$ production and $\Delta$-decay, would yield a ratio of 1.75 for the system 
\nuc{48}{Ca}+\nuc{48}{Ca} and 2 for \nuc{208}{Pb}+\nuc{208}{Pb}. We thus may imagine that the Clebsch-Gordon coefficients for the production may play an important role, but that are not the only ingredient. Furthermore, the constant $Z/A(R)$ for the \rneq option would then yield a pion ratio independent of the centrality. However, we see that  even for \rneq{}, a rise of the pion ratio is visible for peripheral collisions. Additionally it should be stated that the ratios obtained with the \rnlt option are significantly higher than we would expect by the application of a combinatorial model to the participant region taking into account the correct $Z/A$ profile. 

This observation indicates that we have to take into account the rescattering of the pion in nuclear matter. The rescattering is dominated by the absorption of a pion by a nucleon forming a $\Delta$, which afterwards decays again. However, the combination of the different isospin dependent cross sections and of the Clebsch-Gordon coefficients for $\Delta$  production and decay favors the enhancement of the 
$\pi^-/\pi^+$ ratio in neutron-rich matter. This effect becomes even stronger if the outer region of the nucleus (i.e. the region of the last interaction of the pion) becomes neutron-rich, as it is the case e.g. for the \rnlt option. We will come back to this point later.

We can see this effect when decreasing the incident energy to 400 AMeV (left) where the absorption cross sections of pions are higher. We see that for central collisions the ratios for 400 AMeV are higher than those for 1 AGeV and that for the heavy system    
\nuc{208}{Pb}+\nuc{208}{Pb} differences between the \rneq{} and \rnlt{} options emerge even at central collisions. Here the influence of the neutron rich outer skin starts already to show its effects.

\begin{figure*}[htb]
\includegraphics[width=1.4\columnwidth]{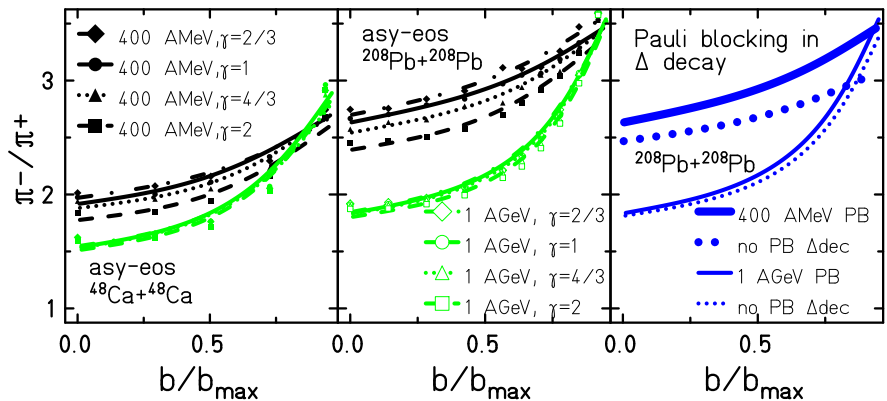}
\caption{The isospin ratio of pions $\pi^-/\pi^+$ as a function of impact parameter $b$ for different parameters of the equation of state of asymetric matter (left \nuc{48}{Ca}, middle \nuc{208}{Pb}) and for different options for the Pauli blocking for the $\Delta$ decay in the system  \nuc{208}{Pb} (right). Green. lines are for  $E_{beam}$=1  AGeV, black lines for  $E_{beam}$=400 AMeV.}
\label{ratios-asyeos}
\end{figure*}

Apart from the effect of rescattering there are other physical processes which have to be discussed. 
A very important question triggering a lot of activity  is that of the influence of the equation of state of asymmetric nuclear matter (asy-eos). It had been suggested that the understanding of the isospin ratios in the FOPI data \cite{Reisdorf:2006ie} should put strong constraints on asy-eos.
To verify this we present in Fig. \ref{ratios-asyeos} calculations with a different stiffness of the density dependence of the asymmetry potential: a small value of the density exponent $\gamma$ denotes a rather soft asy-eos while a large value denotes a rather hard one. 
On the left hand side we see results for the system \nuc{48}{Ca}+\nuc{48}{Ca} while results for \nuc{208}{Pb}+\nuc{208}{Pb} are presented in the mid part. Black lines correspond to an energy of 400 AMeV, while green lines correspond to 1 AGeV incident energy. We see that indeed the asy-eos influences the ratio in central collisions at 400 AMeV while the differences vanish when going to peripheral collisions. This can be understood by realizing that in peripheral collisions a quite small compression is reached, thus the density dependence of the asymmetry potential plays a smaller role, even if the isospin asymmetry is even higher at the surface of the nuclei.
For an incident energy of 1 AGeV the differences vanish for all centralities. Here the effects of creating pions everywhere dominate and small changes due to asymmetry potentials play a minor role.

Finally we want to discuss a further effect, frequently omitted in transport models: The decay of a $\Delta$ may produce a nucleon of moderate energy, especially at low incident energy. Therefore the Pauli blocking of this process may play a significant role.
If now the surrounding region is quite neutron-rich, the Pauli blocking will particularly block the production of neutrons. 
If a $\pi^+$ rescatters in neutron rich matter, there is a high possibility of creating a $\Delta^+$, which decays with the probability 2/3 into $p \pi^0$ (which is a major reason for the influence of rescattering already discussed above) and only by 1/3 into 
$n \pi^+$. The  Pauli blocking thus penalizes that second channel even more and thus enhances the penalty for $\pi^+$. For the rescattering of $\pi^-$ in neutron rich matter the dominant $\Delta^-$ decays always into 
$n \pi^-$. Therefore the Pauli blocking adds no penalty on the $\pi^-$, it only delays its production. This effect has indeed been seen in detail using IQMD.
The right hand side of Fig. \ref{ratios-asyeos} shows this influence of the Pauli blocking for the system 
\nuc{208}{Pb}+\nuc{208}{Pb} at 400 AMeV (thick blue lines) and 1 AGeV (thin blue lines) incident energy. 
Indeed, for $E_{beam}$=400 AMeV  the Pauli blocking of the $\Delta$ decay (full line) enhances the $\pi^-/\pi^+$ ratio significantly 
for all centralities compared to calculations without Pauli blocking (dotted line). For  $E_{beam}$=1 AGeV this effect plays no important role anymore because the available phase space is large.

\section{Conclusion}
In conclusion, we have demonstrated that the centrality dependence of the $\pi^-/\pi^+$ ratio is affected by different physical processes as the density dependence of the asymmetry potential, the Pauli blocking of the $\Delta$-decay and the initial distribution of protons and neutrons in nuclei. While the sensitivity on the neutron skin is most significant in peripheral collisions and vanishes in central collisions at 1 AGeV incident energy, the sensitivity on the density dependence of the asymmetry potential becomes very week at 1 AGeV incident energy but significant at 400 AMeV incident energy. The latter effect is also more prominent in central collisions than in peripheral collisions. The Pauli blocking of the $\Delta$-decay finally has an impact only at lower energies like 400 AMeV, but not at higher energies like 1 AGeV. It becomes more significant at peripheral collisions than in central collisions. These differences may allow to
disentangle the influence of the different processes by systematic studies of the centrality dependence of the $\pi^-/\pi^+$ ratio at different incident energies. The comparison of experimental high precision results with different theoretical models opens the perspective to elucidate at the same time different properties of finite nuclear matter.

\end{document}